\documentclass[pra,aps,amssymb,amsmath,reprint,superscriptaddress,showpacs]{revtex4-1}
\usepackage{bm}
\usepackage{graphicx}
\usepackage{color}

\usepackage{hyperref}
\hypersetup{colorlinks=true}
\usepackage[all]{hypcap} 

\newcommand{\akd}{a^{\dagger}_{k}}
\newcommand{\ak}{a^{\phantom{e\dagger}}_{k}}
\newcommand{\aked}{a^{e\dagger}_{k}}
\newcommand{\ake}{a^{e\phantom{\dagger}}_{k}}
\newcommand{\akod}{a^{o\dagger}_{k}}
\newcommand{\ako}{a^{o\phantom{\dagger}}_{k}}
\newcommand{\w}{\omega}

\newcommand{\ua}{\uparrow}
\newcommand{\da}{\downarrow}
\newcommand{\mr}[1]{\mathrm{#1}}
\newcommand{\ket}[1]{\left| #1 \right\rangle}

\begin{document}

\title{Dynamics of a Qubit in a High-Impedance Transmission Line from a Bath Perspective}

\author{Soumya Bera}
\affiliation{Max-Planck-Institut fuer Physik komplexer Systeme, 01187 Dresden,
Germany}
\affiliation{Institut N\'{e}el, CNRS and Universit\'e Grenoble Alpes, F-38042 Grenoble, France}
\author{Harold U. Baranger}
\affiliation{Department of Physics, Duke University, P. O. Box 90305,
Durham, North Carolina 27708, USA}
\author{Serge Florens}
\affiliation{Institut N\'{e}el, CNRS and Universit\'e Grenoble Alpes, F-38042 Grenoble, France}

\begin{abstract}
We investigate the quantum dynamics of a generic model of light-matter
interaction in the context of high impedance waveguides, focusing on the 
behavior of the photonic states generated in the waveguide. 
The model treated consists simply of a two-level system coupled to a 
bosonic bath (the ohmic spin-boson model). 
Quantum quenches as well as scattering of an incident coherent pulse are 
studied using two complementary methods.
First, we develop an approximate ansatz for the electromagnetic waves 
based on a single multimode coherent state wavefunction;
formally, this approach combines in a single framework ideas from 
adiabatic renormalization, the Born-Markov approximation, and input-output 
theory. 
Second, we present numerically exact results for scattering of a weak 
intensity pulse by using Numerical Renormalization Group (NRG) calculations. 
NRG provides a benchmark for any linear response property throughout the 
ultra-strong coupling regime. 
We find that in a sudden quantum quench, the coherent state approach produces 
physical artifacts, such as improper relaxation to the steady state. 
These previously unnoticed problems are related to the simplified form of the 
ansatz that generates spurious correlations within the bath.
In the scattering problem, NRG is used to find the transmission and reflection 
of a single photon, as well as the inelastic scattering of that single photon. 
Simple analytical formulas are established and tested against the NRG data 
that predict quantitatively the transport coefficients for up to moderate 
environmental impedance.
These formulas resolve pending issues regarding the presence of inelastic losses 
in the spin-boson model near absorption resonances, and could be used for 
comparison to experiments in Josephson waveguide quantum electrodynamics (QED). 
Finally, the scattering results using the coherent state wavefunction approach 
are compared favorably to the NRG results for very weak incident intensity. 
We end our study by presenting results at higher power where the response of 
the system is nonlinear.  
\end{abstract}

\date{\today}

\maketitle

\section{Introduction}

Quantum optics deals with the interaction of matter and electromagnetic waves in 
regimes where the granularity of light comes into play. Typically, 
because both the fine structure constant and atomic dipoles are small, matter-light 
coupling is weak, allowing treatment by simple and controlled methodologies, 
such as the rotating wave approximation (RWA), Born-Markov schemes, and Fock space 
truncation~\cite{Sargent}.  However, recent artificial systems such as superconducting 
waveguides~\cite{Niemczyk,Forn,Astafiev,Abdumalikov,Hoi,vanLoo}, especially 
the ones using Josephson junction elements to boost the circuit impedance~
\cite{Masluk,Bell,Altimiras,Weissl,LeHur}, constitute metamaterials where charge 
density fluctuations mimic an optical-like medium in which the coupling to matter 
(namely superconducting qubits) can be ultra strong. In this situation, a plethora of 
interesting phenomena, such as wide-band frequency conversion~\cite{Goldstein}, large 
non-linearities~\cite{Peropadre}, and non-trivial many-body 
vacua~\cite{Goldstein,Snyman1}, have been theoretically predicted. 

In this context, the standard theoretical approaches mentioned above are insufficient: 
Born-Markov schemes are clearly violated by the large coupling constant,
and counter-rotating terms cannot be neglected anymore. In addition, a
brute force Fock truncation of the full Hilbert space becomes numerically
prohibitive due to the large number of photons generated in the environment, unless 
one can target physical states using an optimal variational basis, such as matrix 
product states (MPS)~\cite{Guo,Zueco,Bruognolo,Chin,PichlerZoler}
or within the systematic coherent state expansion (CSE) pioneered in 
Ref.~\onlinecite{Bera1} and susequently extended to a variety of
dissipative models~\cite{Bera2,ZhaoSubohmic,ZhaoTwobath,Snyman2}.

In this article, we probe 
the idea that multi-mode coherent states 
are a well-adapted tool to deal with ultra-strong coupling quantum optics by
investigating in depth several dynamical properties that are experimentally
relevant, such as population decay, coherence buildup, and photon scattering 
in a large-impedance superconducting waveguide. 
Historically, this approximate single-coherent state approach was 
devised for the ground state of the spin-boson model by Luther and 
Emery~\cite{EmeryLuther} and subsequently by Silbey and Harris~\cite{Silbey,Harris}
and other authors~\cite{ChinSubohmic,Nazir}, before the demonstration that it can be 
turned into a systematic expansion that allows one to reach the exact many-body ground 
state for any coupling strength~\cite{Bera1,Bera2,Snyman2} and an arbitrary bath spectral 
density~\cite{ZhaoSubohmic}.
Starting down this same route, we study quantum dynamics here only at the 
simplest Silbey-Harris level, thus using a single variational coherent state as 
a lowest order approximation for the time-dependent problem (this is also called 
the Davidov ansatz).
Such an approach has been previously applied to study quantum dynamics
for a variety of physical 
problems~\cite{Zhao1,Zhao2,Skrinjar,Zhao3,BeraPV},
and it can be adapted to address scattering of Fock states~\cite{Bermudez}.
The present study allows us to thoroughly assess the merits and drawbacks of this very
economical approach; the numerically-exact full generalization of the systematic 
CSE to the time domain will be addressed in subsequent work.

We will demonstrate here that this simple-minded single coherent state approach
already contains most of the physics at play and naturally bridges from the weak
to ultra-strong coupling regimes. Indeed, population decay is found to
cross over from underdamped to overdamped relaxation at
increasing environmental impedance, as expected physically~\cite{Leggett,Weiss}. 
In addition, this single-coherent state
approximation predicts elastic transmission lineshapes beyond the RWA that in the linear response regime 
match precisely our non-perturbative calculations using the Numerical 
Renormalization Group (NRG). These simulations are
used to derive simple and accurate analytical formulas for
the transmission coefficients and total inelastic deficits, that can then be used to
compare to experiments in waveguide QED.

However, we point out some shortcomings of the single coherent state
dynamics that were not reported so far in the literature. Artifacts are indeed
generically found whenever the two-level system is subject to a strong temporal 
perturbation, such as a quantum quench or a strong irradiation pulse (beyond
the linear response regime). In all these cases, relaxation is found towards
an incorrect steady state where long range correlations are spuriously maintained between the 
two-level system and modes propagating away from it. These artifacts are a consequence 
of the constrained form of the single coherent state ansatz, which neglects
entanglement within the bath states at all times, and are likely to be cured by
extending the present technique to a dynamical version of the systematic coherent state
expansion~\cite{Bera1,Bera2}.

The paper is structured as follows. In Sec.~\ref{Basics} we introduce the
spin-boson model and establish the dynamical equations based on the single
coherent state ansatz. Then, in Sec.~\ref{Quench} we study a simple quantum quench
where the two-level system is subject to the temporal oscillation of a polarization
field. We show that the population decays as physically expected, with a crossover
from underdamped to overdamped behavior at increasing coupling strength.
However, the dynamics of the quantum coherences is incorrect: it builds up 
to a value that does not match the expected steady state. We relate this to
artifacts in the bath dynamics. When the switching of the perturbation is 
made adiabatic, however, proper relaxation is finally recovered, a behavior that we
relate to a factorization property of the emitted wavepacket.
A different physical setup is then considered in Secs.~\ref{ScatteringNRG}
and~\ref{Scattering}, whereupon 
the environment subjects the two-level system to a train of incoming photons.
Transport coefficients are extracted from the quantum dynamics and compared
favorably to NRG calculations. Simple and accurate formulas are also extracted
from these simulations. Some perspectives are given as a conclusion to the paper.

\section{Single coherent state dynamics}
\label{Basics}

\subsection{Model}
We start with the standard spin-boson model~\cite{Leggett,Weiss}, where a
two-level system (for instance, a superconducting qubit in the context of
circuit QED) is coupled to a set of quantized harmonic oscillators describing 
propagating modes in a transmission waveguide. We assume in what follows a
geometry where the qubit is side-coupled to the waveguide, but our results can
be straightforwardly extended to the case of inline coupling. The initial
Hamiltonian reads:
\begin{equation}
H = \frac{\Delta}{2} \sigma_x  - \frac{\sigma_z}{2}\sum_{k \in \mathbb{R}} 
g_k (\ak + \akd) + \sum_{k\in\mathbb{R}} \w_k \akd \ak.
\label{SBinit}
\end{equation}
Here $\Delta$ is the splitting of the two-level system (typically set by 
the Josephson energy associated with the superconducting qubit). We will 
take a linear relation $\w_k = |k|$, where $k$ is the momentum, setting
the plasmon velocity to unity. 
Finally, we parametrize the coupling constant as $g_k = \sqrt{2 \alpha |k|}
\theta(\w_c-\w_k)$.
These assumptions accurately describe superconducting waveguides at
characteristic energies that are well below the cutoff frequency $\w_c$.
As a result, the spectral density in the continuum limit reads 
$ J(\w) = \sum_{k>0} \pi g_k^2 \delta(\w-\w_k)
= 2 \pi \alpha\, \w\,  \theta(\w_c-\w)$,
where, in case of electric coupling, the dimensionless coupling strength 
$\alpha$ is proportional to the impedance of the waveguide.
As is standard practice, it is useful to fold the problem onto a half-line, by 
defining even and odd modes,
\begin{equation}
a_{k}^{e}=\frac{1}{\sqrt{2}}\left(a_{k}+a_{-k}\right)\quad\textrm{and}\quad
a_{k}^{o}=\frac{1}{\sqrt{2}}\left(a_{k}-a_{-k}\right),
\end{equation}
so that the Hamiltonian~(\ref{SBinit}) can be rewritten:
\begin{equation}
H = \frac{\Delta}{2} \sigma_x  - \frac{\sigma_z}{2}\sum_{k>0}
g_k (\ake + \aked) + \sum_{k>0} \w_k [\aked \ake\!+\akod \ako].
\label{SB}
\end{equation}

\subsection{Dynamical ansatz and quantum equations of motion}

In the small impedance limit, $\alpha\ll1$, it is customary to invoke
the rotating wave approximation (RWA)~\cite{Sargent}, where Hamiltonian~(\ref{SB})
is truncated such that qubit levels dressed only by adjacent Fock states 
are included.
However, this approximation breaks down already for $\alpha\gtrsim0.1$,
and so for our purposes the model must be addressed in its full complexity. 
A particular defect of the RWA is the lack of 
many-body renormalization---that is,
the strong reduction of the bare tunneling energy
$\Delta$ to a smaller value $\Delta_R$. This renormalization effect is, however, 
well-described by an alternative approach, where
the dressing of the qubit levels occurs via coherent 
states~\cite{EmeryLuther,Silbey,Harris,Nazir,Bera1}.
At the lowest degree of approximation, a single multimode coherent state 
is introduced for each qubit state, so that the time-dependent state vector
takes the form of the following simple ansatz:
\begin{eqnarray}
\label{Psi}
\ket{\Psi(t)} &=& p(t) \ket{\ua} e^{\sum_k [ f_k(t) \aked - f_k^*(t)\ake]}
\ket{0}\\
\nonumber
&& + q(t) \ket{\da} e^{\sum_k [h_k(t) \aked - h_h^*(t)\ake] } \ket{0}.
\end{eqnarray}
Here $p(t)$ and $q(t)$ are the time-dependent amplitudes of the dressed
qubit states (and map the entire Bloch sphere), while $f_k(t)$ and $h_k(t)$ 
denote complex displacements of the associated bath oscillators dressing the qubit. 
The bath states assume thus a simplified form, where entanglement between the
various modes is simply neglected. As a particular case, note that
the approximate Silbey-Harris state that is obtained for the ground state of
Hamiltonian~(\ref{SB}) satisfies the relations 
(obtained by energy minimization)
$p=-q=1/\sqrt{2}$, $f_k=-h_k\equiv f_k^\mr{SH}=(1/2) g_k/(\w_k+\Delta_R)$, 
where $\Delta_R$ is the renormalized qubit tunnel amplitude, an important 
parameter in what follows.

Only the even modes of the waveguide appear in the ansatz (\ref{Psi}) 
as the odd modes are decoupled from the qubit. 
In Section~\ref{Quench}, the odd modes will be taken in their 
vacuum state and will not be considered in the dynamics. However, the transport 
conditions considered in Sec.~\ref{Scattering} will require inclusion of the odd modes, 
which is trivially done since their evolution is given by the 
free part of Hamiltonian~(\ref{SB}).

The quantum dynamics will be piloted by the real Langrangian density 
$\mathcal{L}= \big<\Psi(t)|\frac{i}{2} \overrightarrow{\partial_t} 
-\frac{i}{2} \overleftarrow{\partial_t} 
- \mathcal{H}|\Psi(t)\big>$, with Euler-Lagrange type of equations of motion,
as resulting from the Dirac-Frenkel time-dependent variational 
principle~\cite{Saraceno}:
\begin{eqnarray}
\nonumber
\frac{d}{dt} \frac{\partial \mathcal{L}}{\partial \dot{f}_k} =
\frac{\partial \mathcal{L}}{\partial f_k}, \;
\frac{d}{dt} \frac{\partial \mathcal{L}}{\partial \dot{h}_k} =
\frac{\partial \mathcal{L}}{\partial h_k},\;
\frac{d}{dt} \frac{\partial \mathcal{L}}{\partial \dot{p}} =
\frac{\partial \mathcal{L}}{\partial p}, \;
\frac{d}{dt} \frac{\partial \mathcal{L}}{\partial \dot{q}} =
\frac{\partial \mathcal{L}}{\partial q}.\\
\end{eqnarray}
This results in the set of dynamical equations:
\begin{eqnarray}
\label{eqf}
i \dot{f}_k &=& \w_k f_k - \frac{g_k}{2}
-\frac{q \Delta}{2p} (f_k-h_k) \left\langle f|h\right\rangle,\\
\label{eqh}
i \dot{h}_k &=& \w_k h_k + \frac{g_k}{2}
-\frac{p \Delta}{2q} (h_k-f_k) \left\langle h|f\right\rangle,\\
\nonumber
i\dot{p} &=& \frac{\Delta}{2}q \left\langle f|h\right\rangle
-p \sum_k\frac{g_k}{4}(f_k+f_k^*)+p\sum_k \w_k |f_k|^2\\
\label{eqp}
&& -i\frac{p}{2}\sum[\dot{f}_k f_k^* - \dot{f}_k^* f_k],\\
\nonumber
i\dot{q} &=& \frac{\Delta}{2}p \left\langle h|f\right\rangle
+q \sum_k\frac{g_k}{4}(h_k+h_k^*)+q\sum_k \w_k |h_k|^2\\
\label{eqq}
&& -i\frac{q}{2}\sum[\dot{h}_k h_k^* - \dot{h}_k^* h_k],
\end{eqnarray}
where $\left\langle f|h\right\rangle =
e^{\sum_k[f_k^*h_h-|f_k|^2/2-|h_k|^2/2]}$.
In practice, Eqs.~(\ref{eqf}-\ref{eqh}) are substituted into
Eqs.~(\ref{eqp}-\ref{eqq}), so that an independent set of first-order
non-linear differential equations is obtained, which can be
efficiently solved by standard Runge-Kutta techniques, with a linear 
scaling of the computational cost in the number of bosonic modes.

Such dynamical equations have a long history from polaron
physics~\cite{Skrinjar,Zhao3,BeraPV} to dissipative quantum
mechanics~\cite{Zhao1,Zhao2}, and have been previously derived
in the literature. Our purpose in this paper is to benchmark carefully
the physical results that they lead to, in order to pinpoint the
advantages and drawbacks in using them in the specific context of
waveguide QED. We first consider in the next section the situation 
of quantum quenches, before turning to the investigation of scattering
properties.

\section{Population decay and coherence buildup}
\label{Quench}

\subsection{Sudden quantum quench}
\label{Sudden}

We investigate here a standard protocol, where the qubit is initialized at time
$t=0$ in the state $\ket{\ua}$, while the bath is taken in the vacuum $\ket{0}$.
(Note that $\ket{\ua}$ is not the excited state of the qubit but rather a
quantum superposition of the ground and excited state.)
The qubit then evolves at later times according to the full spin-boson
Hamiltonian~(\ref{SB}), and progressively relaxes to its many-body ground state while
energy is dissipated into the bath. This theoretical problem has been considered
by a great variety of methods, such as quantum Monte Carlo~\cite{Egger,Ankerhold}, stochastic 
expansions~\cite{Gisin,Grabert,Orth}, time-dependent NRG~\cite{BullaAnders,Hofstetter}, 
systematic variational dynamics~\cite{WangThoss},
and analytical weak-coupling
calculations~\cite{Schoeller1,Kehrein,Kennes,Schoeller2}, but mostly from the
perspective of the qubit dynamics. Indeed, in addition to the relaxation properties 
of the qubit itself, we will examine here in depth the behavior of the states in the 
bath. This joint qubit and bath dynamics is approximated using the simple 
equations of motion Eqs.~(\ref{eqf})-(\ref{eqq}), with the initial conditions 
$p=1$, $q=0$, $f_k=h_k=0$. In fact, for numerical stability reasons, one must
give a small non-zero value, typically $q=10^{-3}$ at initial times.

By symmetry, the two-level system shows no polarization along the 
$z$-axis in its ground state (this applies only in the delocalized phase
$\alpha<1$, otherwise spontaneous polarization does occur).
Due to the presence of the transverse field $\Delta$, 
one expects precession of the spin and so damped oscillation
of $P(t)\equiv\left<\sigma_z(t)\right>$ reaching zero in the long-time limit.
In addition, it is known that the dynamics crosses over from an underdamped
to an overdamped regime as dissipation reaches values around $\alpha=0.5$.
Finally, one expects the oscillations to occur at a renormalized
frequency $\Delta_R \simeq \Delta (\Delta/\w_c)^{\alpha/(1-\alpha)}$ 
with a damping rate $\Gamma \simeq \alpha \Delta_R$.

\begin{figure}[th]
\includegraphics[width=1.0\linewidth]{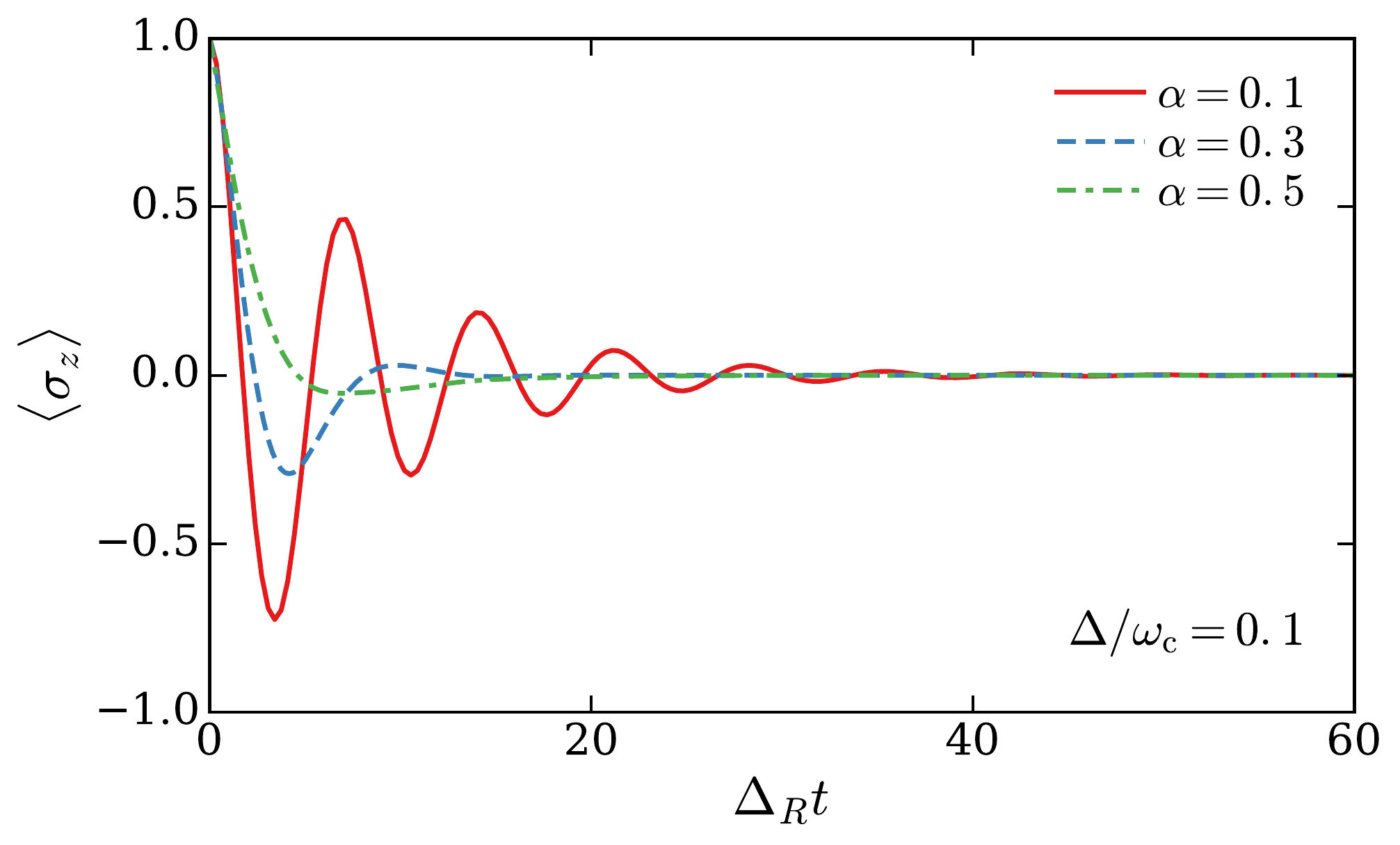}
\includegraphics[width=1.0\linewidth]{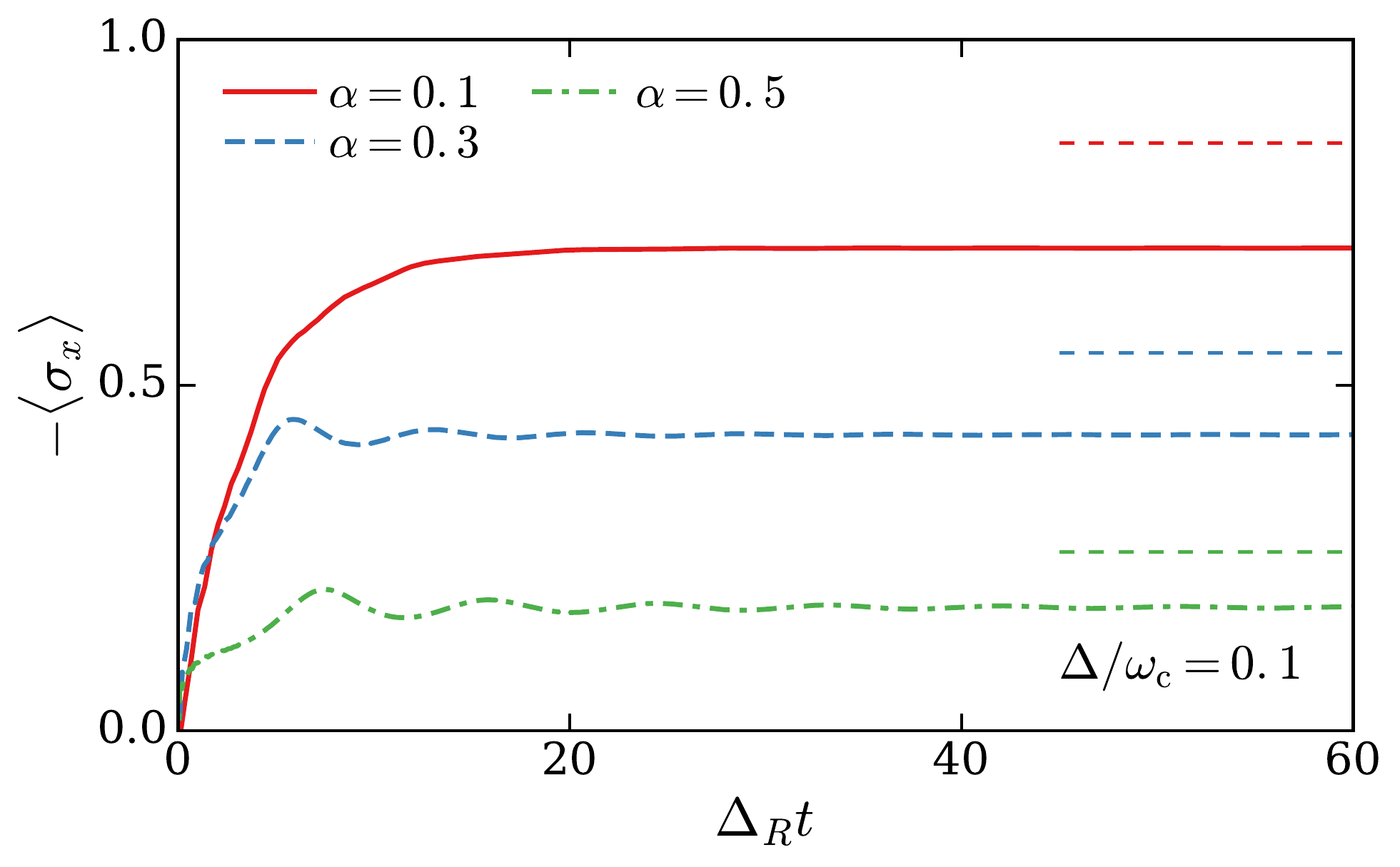}
\caption{(Color online) Population decay $\big<\sigma_z(t)\big>$ (upper panel) and 
coherence buildup $\big<\sigma_x(t)\big>$ (lower panel) of the two-level system 
for several values of dissipation 
$\alpha=0.1,0.3,0.5$ at $\Delta/\omega_c=0.1$. The dashed lines on the vertical 
axis of the lower panel denote the values for the ground state coherence that are expected from the 
static single-coherent state approximation. The discrepancy with the long time
limit of $\big<\sigma_x\big>$ originates in  
spurious correlations between the bath and the two-level system.} 
\label{QubitDynamics}
\end{figure}

One remarkable achievement of the single-coherent-state dynamics is that all 
these non-trivial features of the sudden quench dynamics are
qualitatively obtained, as can be seen from the top panel in Fig.~\ref{QubitDynamics}.
Comparison to the existing literature shows however that the precise form of the population decay
obtained from this single coherent-state approximation is not fully accurate. 
First, the exact renormalized qubit splitting differs from the single coherent state 
result by numerical factors that can be sizable, as was shown recently by extensive 
CSE calculations in the ground state~\cite{Bera2,Snyman2}. Second, we see that
for $\alpha=0.5$, the qubit is still slightly underdamped, while it is
known~\cite{Kennes} that the dynamics should be strictly overdamped, without any  sign change in $P(t)$.

\begin{figure}[th]
\includegraphics[width=1.0\linewidth]{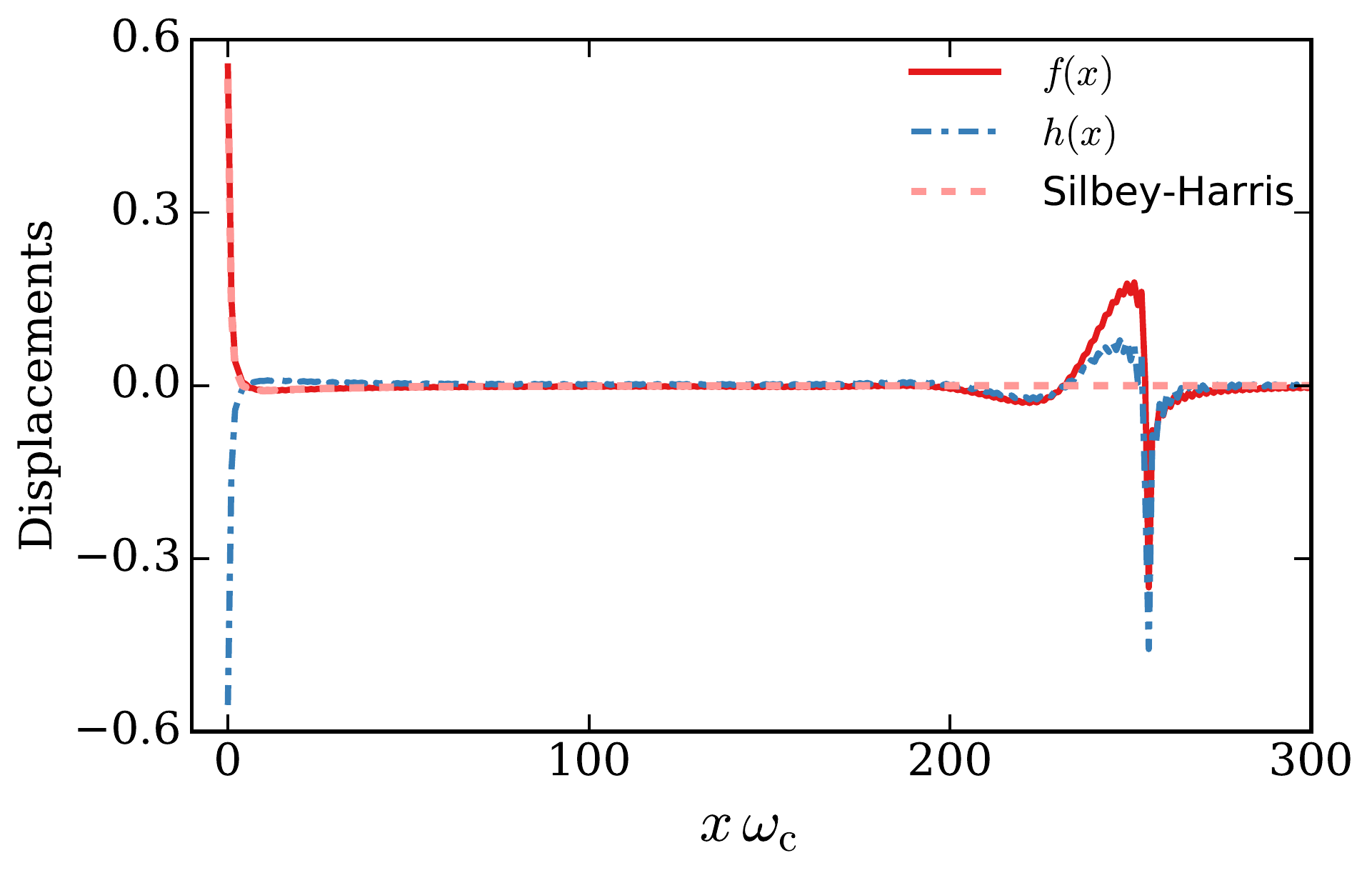}
\caption{(Color online) Real space displacements $f(x)$ and $h(x)$ associated with 
the $\ua$ and $\da$ components of the wavefunction, respectively, shown for a
long time after the instantaneous quench process discussed in the text
(parameters are $\Delta=0.1$ and $\alpha=0.2$).
At short distances, the Silbey-Harris ground state is stabilized, while at large 
distances a wavepacket is propagating away from the qubit located at the origin. The
lack of factorization, $f(x)\neq g(x)$ for $x>220$, is associated with spurious
correlations between the emitted wavepacket and the two-level system, that are
responsible for the improper relaxation seen in Fig.~\ref{QubitDynamics}.
}
\label{QuenchBath}
\end{figure}

However, a more problematic and qualitative issue arises when monitoring the qubit coherence
$\left<\sigma_x(t)\right>$, which was not reported in previous 
studies~\cite{Zhao1,Zhao2}.
The lower panel in Fig.~\ref{QubitDynamics} shows that the qubit coherence builds as expected 
qualitatively, but that the long-time limit is in stark disagreement with the 
value found from the single-coherent-state Silbey-Harris approximation in the ground 
state~\cite{Silbey,Nazir,Bera1}.
Note that the single coherent state ansatz~(\ref{Psi}) captures very precisely the 
ground state for $\alpha<0.2$, so the discrepancy in the long time
value of $\left<\sigma_x(t)\right>$ is unexpected.
The origin of the problem lies in the states of the bath that carry
energy away from the qubit.
By monitoring the bath displacements $f_k$ and $h_k$ in real
space (as obtained by Fourier transform), one can see in Fig.~\ref{QuenchBath} 
that for distances $x\lesssim 1/\Delta_R$, an entanglement cloud~\cite{Snyman1} 
forms between the qubit and the waveguide, which maps perfectly onto the Silbey-Harris
predictions for the ground state (dotted line). This confirms that the origin of 
the improper relaxation must lie in the propagating waves, that are seen in
the plot at larger distances. This can be understood physically in the limit 
$\alpha\ll1$ as follows. Let us define the bare qubit ground and excited eigenstates 
$\ket{g}=(\ket{\ua}-\ket{\da})/\sqrt{2}$
and $\ket{e}=(\ket{\ua}+\ket{\da})/\sqrt{2}$. Our initial state reads:
$\ket{\Psi(t=0)} = \ket{\ua}\otimes\ket{0} = (1/\sqrt{2})\ket{g}\otimes\ket{0}
+ (1/\sqrt{2})\ket{e}\otimes\ket{0}$. At weak dissipation, the configuration
$\ket{g}\otimes\ket{0}$ is close to the actual ground state, and
does dot experience any time evolution. In contrast, the excited state
$\ket{e}\otimes\ket{0}$ will decay towards the low energy state of the
qubit~\cite{Sargent} while emitting a single photon at the resonant frequency $\w=\Delta$.
Thus the complete wavefunction in the long time limit should read:
\begin{eqnarray}
\ket{\Psi(t=\infty)} &=& \frac{1}{\sqrt{2}}\ket{g}\otimes\ket{0} + 
\frac{1}{\sqrt{2}}\ket{g}\otimes a^\dagger_\Delta \ket{0} \\
&=& \frac{(\ket{\ua}-\ket{\da})}{\sqrt{2}} \otimes 
\frac{\ket{0}+a^\dagger_\Delta\ket{0}}{\sqrt{2}}.
\end{eqnarray}
Clearly the qubit states and the emitted photons are unentangled
in the long time limit, which is in contrast to the outcome of the single
coherent state approximation. Indeed, from Fig.~\ref{QuenchBath}, one
sees that the real space displacements $f(x)$ and $h(x)$ are close to
each other, but not strictly equal. This generates some spurious correlations
with the qubit states, hence the improper value for the long-time coherence.

\subsection{Adiabatic switching}

\begin{figure}[tb]
\includegraphics[width=1.0\linewidth]{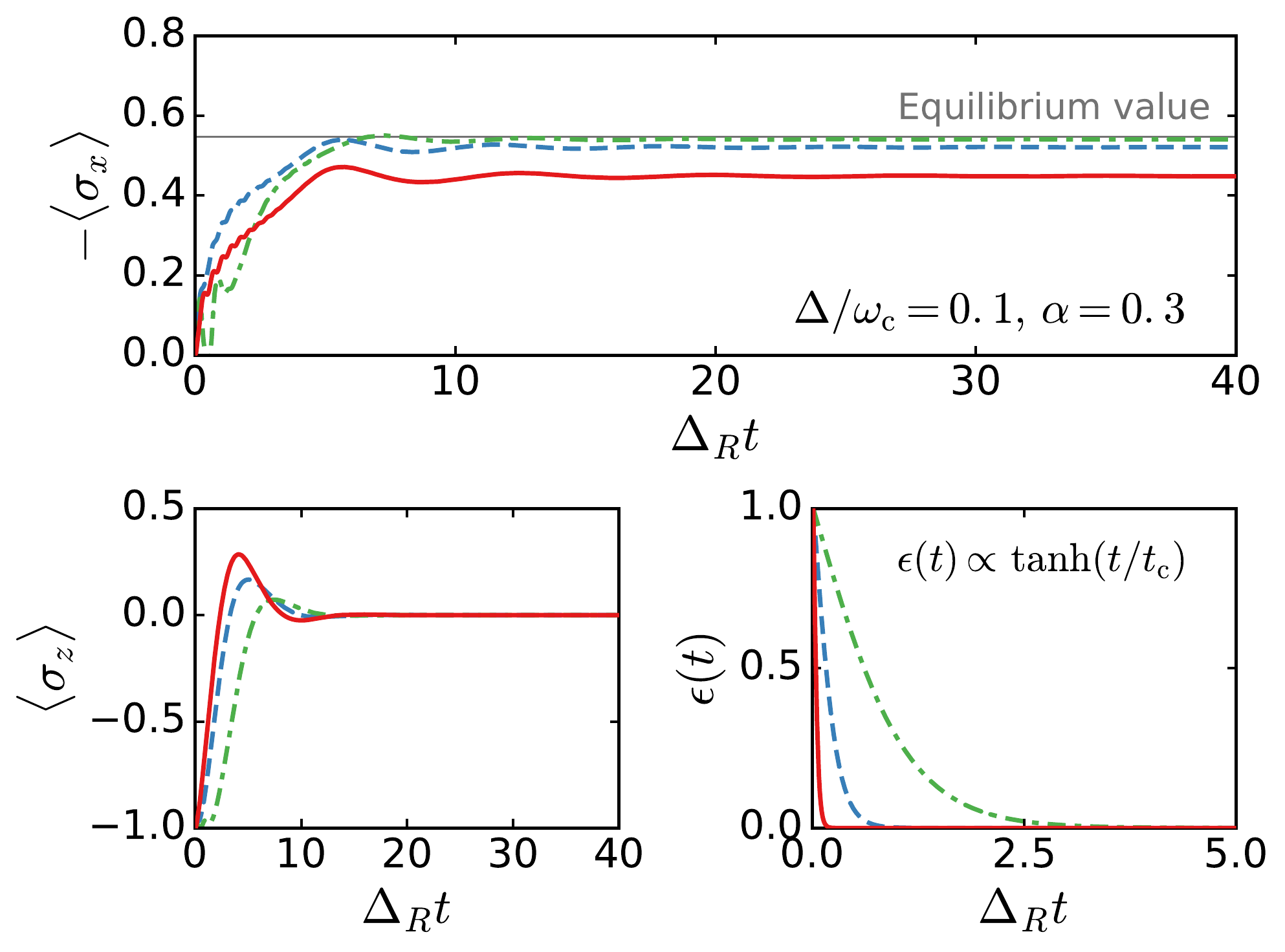}
\includegraphics[width=1.0\linewidth]{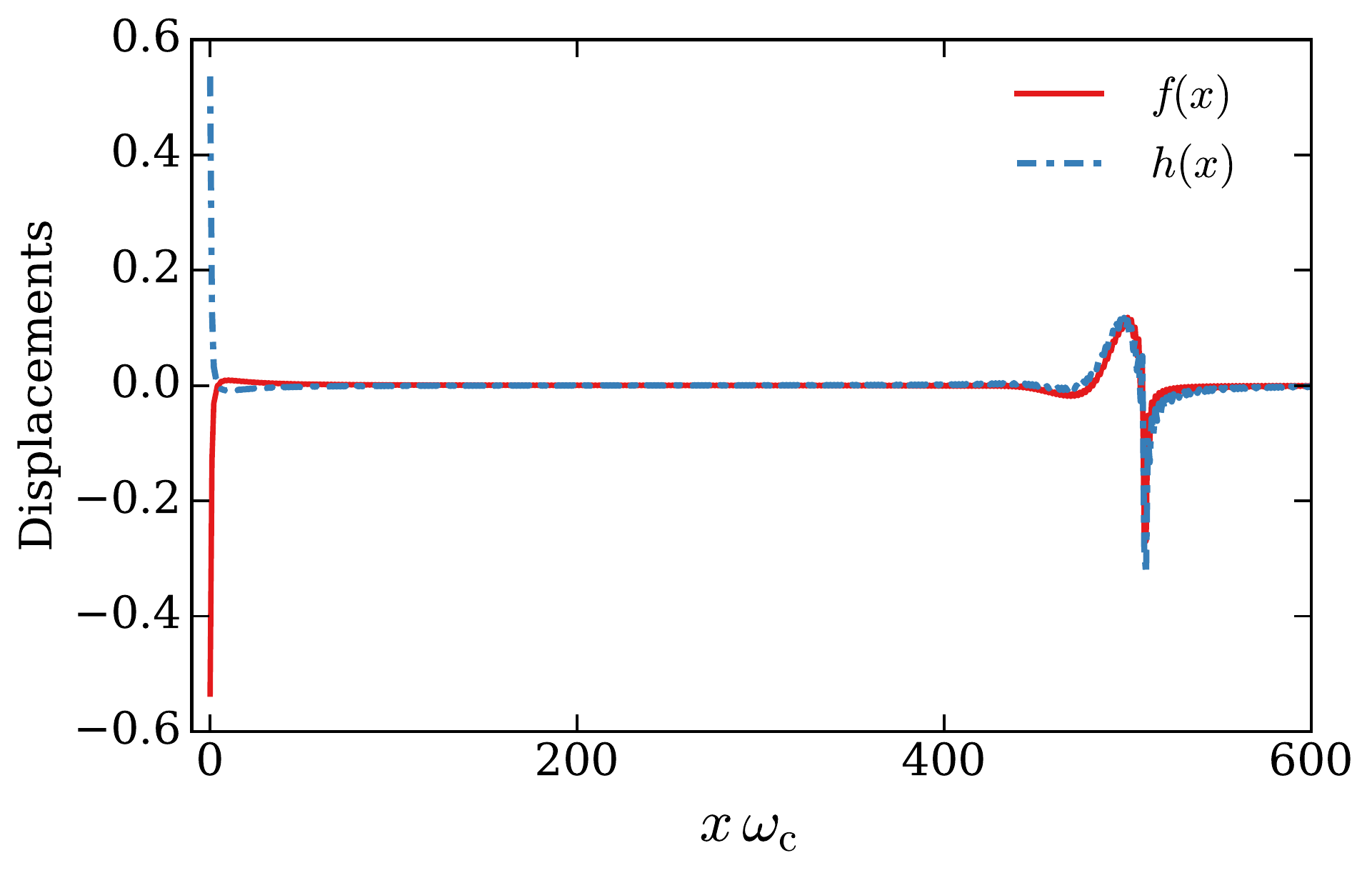}
\caption{(Color online) Top panel: Coherence $\big<\sigma_x(t)\big>$ as a function of time
for three different switching processes (from sudden to adiabatic), that are given by the temporal 
profile of the local field shown in the rightmost middle panel. The saturation to the 
equilibrium value is obtained for an adiabatic switching, while the sudden quench shows 
improper relaxation. The leftmost middle panels shows the associated population decay 
$\big<\sigma_z(t)\big>$.
Bottom panel: real space displacement $f(x)$ and $h(x)$, associated to
the $\ua$ and $\da$ components of the wavefunction respectively, shown for a
long time after the adiabatic quench. Now factorization is correctly recovered,
in agreement with the proper relaxation value of $\big<\sigma_x(t)\big>$ at long
times.}
\label{AdiabaticQuench}
\end{figure}

These moderate artifacts, which are found in previous
out-of-equilibrium NRG calculations as well~\cite{Nghiem}, are related to the 
sudden form of
the quench. Indeed, if the qubit is driven adiabatically, the correct relaxation 
occurs within the single coherent-state scheme. 
To show this, we still subject the qubit to a time-dependent polarization 
field $\epsilon(t)$ along the $z$-axis, but now switch it off gradually.
The resulting coherence
$\left<\sigma_x(t)\right>$ is shown in Fig.~\ref{AdiabaticQuench} for three 
values of the switching time.

For a short switching time, the
long-time value of the coherence is underestimated, as for the sudden quenches. 
However, for a more adiabatic pulse, good convergence towards the ground state value is recovered. At the same time,
one can check in Fig.~\ref{AdiabaticQuench} that the emitted wavepacket is factorized with respect to the short distance cloud: indeed, the real space
displacements $f(x)$ and $h(x)$ are now equal to each other at large distances,
ensuring proper factorization.

These observations thus show the merits and drawbacks of the popular
single-coherent-state dynamics, also known as the Davidov dynamical
ansatz~\cite{Zhao1,Zhao2}. After a sudden quench, the polarization dynamics and
final wavefunction near the origin are captured reasonably well, but the
coherence and eventual disentanglement of the qubit and the traveling photon are
not. For a sufficiently adiabatic quench, all quantities are correctly captured 
by the single-coherent-state dynamics.

\section{Scattering properties: Numerical renormalization group}
\label{ScatteringNRG}

\begin{figure}[b]
\includegraphics[width=1.0\linewidth]{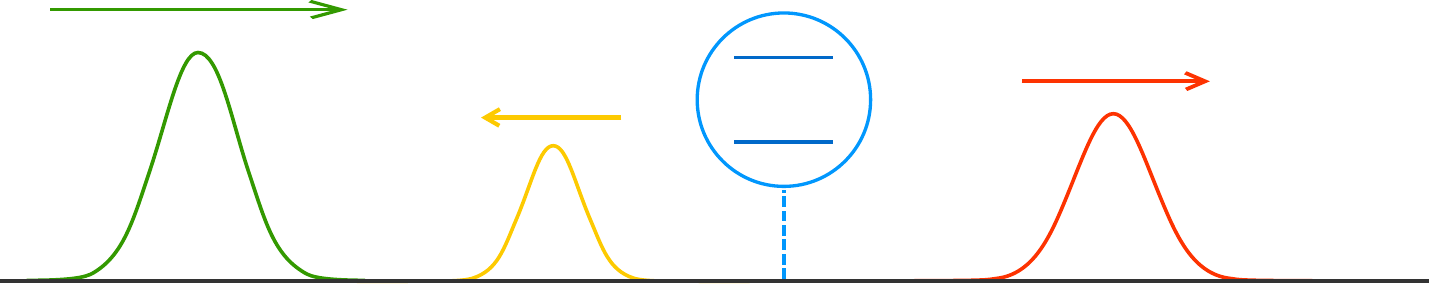}
\caption{(Color online) Waveguide setup considered here, where a qubit is
side-coupled in a two-terminal geometry to a photonic waveguide. Transport 
properties are described by scattering processes depicted by a set of 
ingoing and outgoing wavepackets.}
\label{waveguide}
\end{figure}

In this second part of the paper, we wish to study a typical situation
in quantum optics in the specific context of waveguide QED: we investigate
the scattering properties of photons in the two-terminal setup depicted
in Fig.~\ref{waveguide}, but in a domain of ultra-strong coupling. 
This question was investigated by many authors in a regime of weaker coupling,
$\alpha\ll1$, using for instance the RWA~\cite{Shen,Zheng,Wang} or extensions 
to scattering of the Wigner-Weisskopf theory~\cite{Chen} (based on Fock state 
truncation). In order to tackle the case of large $\alpha$, we first develop a 
numerically exact approach to the photon scattering properties in the weak 
intensity limit. This is
done by using the numerical renormalization group method~\cite{Bulla,Freyn}. The
results provide an important benchmark for the coherent state dynamics treated
in the next section and, indeed, 
for any calculations of scattering in waveguide QED.

\subsection{Elastic and inelastic transport coefficients from the numerical
renormalization group}
 
We start by defining the real-time retarded equilibrium spin susceptibility:
\begin{equation}
\chi(t) = -\frac{i}{4} \theta(t) \left<\mr{GS}|[\sigma_z(t),\sigma_z(0)]|\mr{GS}\right>
\end{equation}
with $\left|\mr{GS}\right>$ the full many-body ground state (note that $\chi(t)$
is a purely real function). Inserting a complete eigenbasis of states
$\{\ket{a}\}$ with respective energies $E_a$, one readily obtains:
\begin{equation}
\chi(t) = - \frac{1}{2} \theta(t) \sum_a |\left<\mr{GS}|\sigma_z|a\right>|^2
\sin[(E_a-E_\mr{GS})t].
\end{equation}
\begin{figure}[tb]
\begin{flushleft}
\includegraphics[width=0.99\columnwidth]{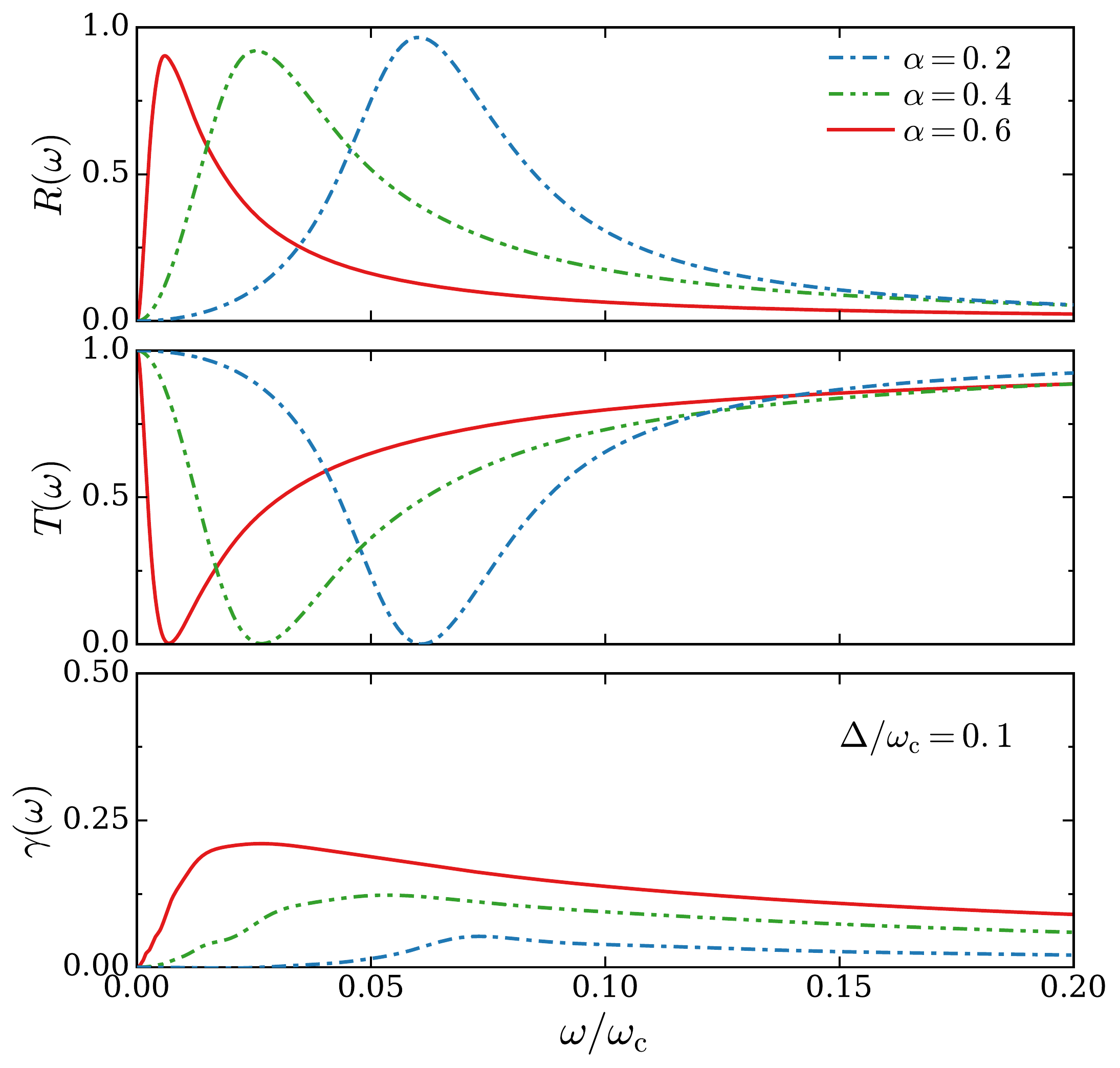}
\end{flushleft}
\caption{(Color online) Reflection coefficient (top panel), transmission
coefficient (middle panel) and total inelastic losses (bottom panel) from
the NRG calculations at three values of dissipation $\alpha=0.2,0.4,0.6$ for
$\Delta=0.1$. The curves with largest $\alpha$ show a resonant peak at the
smallest value of the renormalized qubit splitting $\Delta_R$.}
\label{NRGdata}
\end{figure}
This leads to the frequency-resolved spin susceptibility
(decomposed into real and imaginary parts):
\begin{equation}
\chi(\w) \equiv \chi'(\w)+i\chi''(\w) = \int_{-\infty}^{+\infty} \mr{d}t \;
e^{i\w t} \chi(t).
\end{equation}
In particular, in terms of the Lehman spectral
decomposition onto the complete eigenbasis, the imaginary part reads:
\begin{eqnarray}
\chi''(\w) &=& \frac{\pi}{4} \sum_a |\left<\mr{GS}|\sigma_z|a\right>|^2 \left[
\delta(\w+E_a-E_\mr{GS})\right. \\
&& - \left. \delta(\w-E_a+E_\mr{GS})\right].
\end{eqnarray}
One thus obtains from the above decomposition a sum rule that will 
be important in what follows:
\begin{equation}
\label{sumrule}
\int_{-\infty}^{+\infty} \!\! \mr{d}\w \; \chi''(\w) \, \mr{Sign}(\w) = \frac{\pi}{2}.
\end{equation}

Now, from linear-response theory~\cite{LeHur,Goldstein} and using an
exact identity 
for the Green funciton of the bosonic modes,
\begin{eqnarray}
G_{kk'}(\w) & \equiv & \big<a_k^\dagger(\w) a_{k'}^{\phantom{\dagger}}(-\w)\big>\nonumber \\
 &=& \frac{\delta_{kk'}}{\w-\w_k} + \frac{g_k g_{k'} \chi(\w)}
{(\w-\w_k)(\w-\w_{k'})}
\end{eqnarray}
which relates the scattering matrix to the qubit response, one obtains the
reflection and transmission coefficients,
\begin{eqnarray}
\label{RKubo}
R(\w) & = & (2\pi \alpha \w)^2 |\chi(\w)|^2\\
\label{TKubo}
T(\w) & = & (2\pi \alpha \w)^2 [\chi'(\w)]^2+[1-2\pi\alpha\w\chi''(\w)]^2 ,
\end{eqnarray}
as well as the total inelastic deficit,
\begin{equation}
\gamma(\w) \equiv 1-R(\w)-T(\w) =  4\pi \alpha \w\chi''(\w)
-2(2\pi \alpha \w)^2 |\chi(\w)|^2.
\end{equation}
As they are derived from linear response theory, the reflection and transmission probabilities here are those of a single incoming photon.

These quantities are shown in Fig.~\ref{NRGdata} for increasing values
of dissipation.
The reflection/transmission coefficient shows as expected a peak/dip
at the qubit absorption frequency $\Delta_R$,
which progressively renormalizes to smaller values as $\alpha$ increases. 
At the same time, one notes that the peak value in $R(\w)$ is slightly lower
than unity, with a deviation that increases with enhanced dissipation. This is
in contrast to the approximate results of Ref.~\onlinecite{LeHur} but in agreement
with exact calculations at the Toulouse point~\cite{Bulla,Goldstein}, which
read:
\begin{equation}
\chi_\mathrm{Toulouse}(\w) = \frac{1}{2\pi\w} \frac{1}{\w+i\Gamma}
\log(1-2 i \w).
\end{equation}
As a consequence, the inelastic deficit becomes more and more important as $\alpha$
grows, due to stronger photon non-linearities, as shown in the bottom panel.
One notes in particular that the deficit $\gamma(\w)$ peaks above $\Delta_R$
and shows very long tails, which are reminiscent of 
the inelastic contribution to scattering for fermionic Kondo 
impurities~\cite{Fritz}, although the physical quantities do not correspond
strictly here. As previously noted by Goldstein {\it et al.}~\cite{Goldstein}, the
total inelastic losses are quite important at strong dissipation, and reach above
20$\%$ at $\alpha=0.6$. We note in addition a surprising result from 
the NRG: while the reflection deficit is sizable [for large impedance, typically up 
to 10\% deviation
from unitary scattering at the peak value of $R(\w)$], 
the transmission deficit is very small---the transmission dip goes to a tiny 
(yet non-zero) value.
It is practically not possible to see this small background in the middle panel of
Fig.~\ref{NRGdata}, but one can verify from the exact Toulouse formula at $\alpha=0.5$
that the minimal transmission is of order $10^{-3}$, a very small number
to which we cannot give a simple physical interpretation at this stage.

\subsection{Analytical comparisons}
We provide here some analytical insights, comparing previously derived theories 
to our numerical simulations. We also derive new phenomenological formulas
that match the NRG results quite well, and that could be used in practice 
for easier comparisons to experiments. We do not include checks to the
exact Toulouse limit~\cite{Goldstein}, as it has been shown
earlier~\cite{Bulla} that the NRG reproduces the spin susceptibility quite
accurately at $\alpha=0.5$.

\begin{figure}[tb]
\includegraphics[width=0.99\columnwidth]{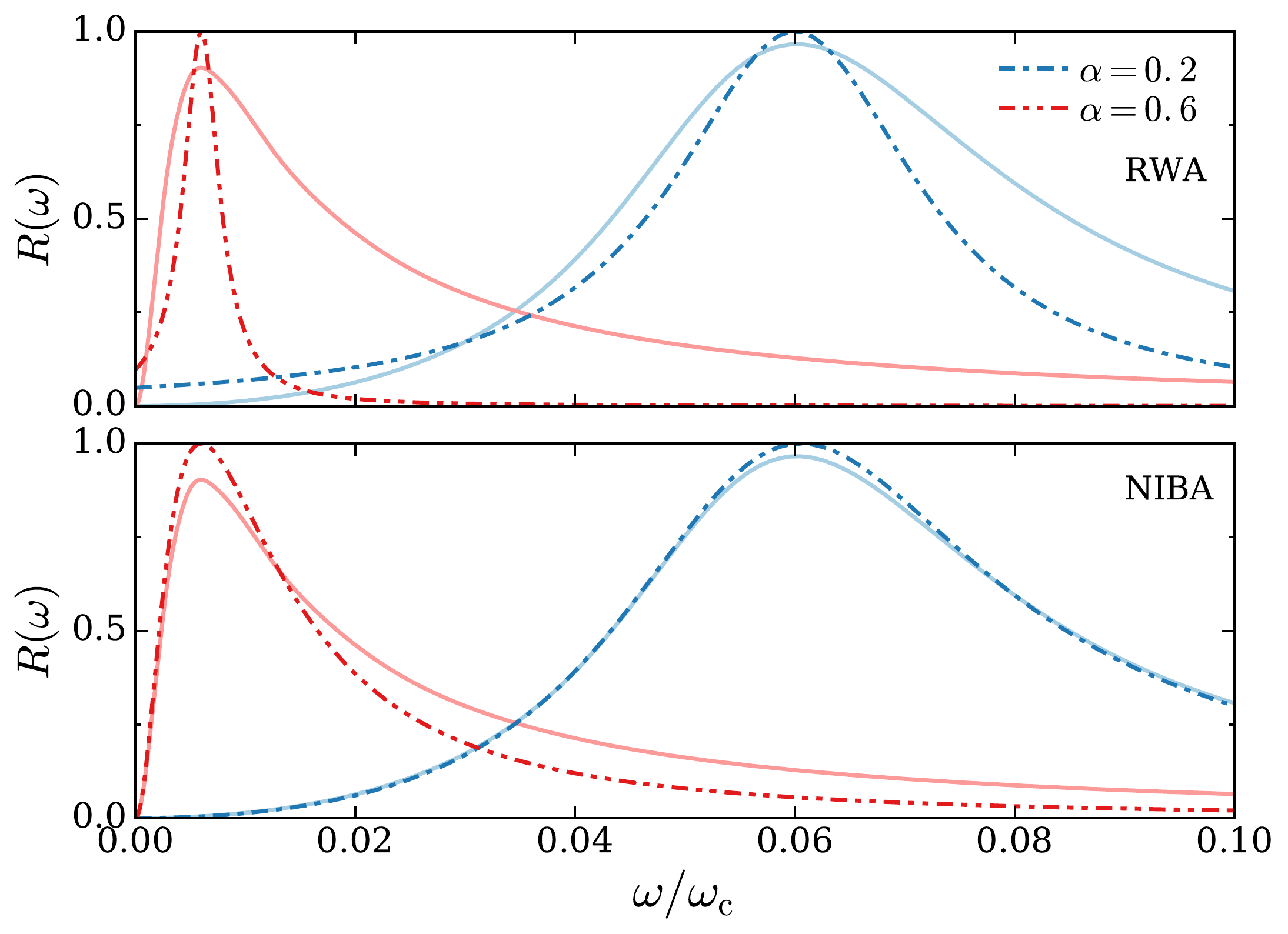}
\caption{(Color online) Reflection coefficient comparing the RWA Lorentzian 
lineshape Eq.~(\ref{RRWA}) (dashed lines, upper panel) and the NIBA lineshape
Eq.~(\ref{RNIBA}) of Ref.~\onlinecite{LeHur} (dashed lines, lower panel)
to the exact NRG calculations (full lines) at two values of dissipation 
$\alpha=0.2,0.6$ for $\Delta=0.1$. An effective
renormalized qubit splitting $\Delta_R$ and a renormalized linewidth $\Gamma$
were used as fitting parameters within the RWA and NIBA formulas, for better 
comparison of the actual lineshapes.}
\label{CompareRWA}
\end{figure}

First, we compare the NRG data to approximate results based on the rotating 
wave approximation used routinely in quantum optics~\cite{Sargent}, which leads 
to the following $t$-matrix for a unidimensional waveguide:
\begin{equation}
t_\mr{RWA}(\w) = \frac{\w-\Delta}{\w-\Delta+i\Gamma}.
\end{equation}
In this approximation, the qubit level is not renormalized, but acquires
a lifetime, with a rate $\Gamma\propto\alpha\Delta$. This leads to the
reflection and transmission coefficients:
\begin{eqnarray}
\label{RRWA}
\!\!\!\!\!\!R_\mr{RWA}(\w) &=& |1-t_\mr{RWA}(\w)|^2 =
\frac{\Gamma^2}{(\w-\Delta)^2+\Gamma^2}\\
\!\!\!\!\!\!T_\mr{RWA}(\w) &=& |t_\mr{RWA}(\w)|^2 =
\frac{(\w-\Delta)^2}{(\w-\Delta)^2+\Gamma^2}.
\end{eqnarray}
Clearly, scattering is unitary in the RWA, since $R_\mr{RWA}(\w)=1-T_\mr{RWA}(\w)$. 
Due to the lack of inelastic effects, the RWA lineshape 
should become less accurate as dissipation increases. We
see indeed sizable deviations when $\alpha>0.1$ in the upper panel of Fig.~\ref{CompareRWA} 
by a comparison to NRG for the reflection coefficient. Note that we allow 
here as fitting parameters for the RWA lineshape the renormalized qubit 
frequency and linewidth, which improves the agreement somewhat artificially.

This defect of the RWA has motivated improved perturbative
calculations by K.\ Le~Hur~\cite{LeHur}, based on an approximate form for the spin
susceptibility that derives from analogy to the Non Interacting Blip Approximation
(NIBA)~\cite{Leggett}:
\begin{equation}
\label{chiNIBA}
\chi_\mr{NIBA}(\w) = \frac{\Delta_R}{\Delta_R^2-\w^2 - i 2\pi\alpha\w\Delta_R}.
\end{equation}
   From Eqs.~(\ref{RKubo})-(\ref{TKubo}), 
this yields an approximate form of the 
reflection and transmission coefficients,
\begin{eqnarray}
\label{RNIBA}
R_\mr{NIBA}(\w) &=& 
\frac{(2\Gamma\w)^2}{(\Delta_R^2-\w^2)^2+(2\Gamma\w)^2}\\
T_\mr{NIBA}(\w) &=& 
\frac{(\Delta_R^2-\w^2)^2}{(\Delta_R^2-\w^2)^2+(2\Gamma\w)^2},
\end{eqnarray}
from which the RWA expression~(\ref{RRWA}) is recovered in the limit $\Gamma\ll\Delta_R$.
These formulas can be seen, however, to again satisfy $R_\mr{NIBA}(\w)+T_\mr{NIBA}(\w)=1$
and thus miss inelastic losses, as pointed out in Ref.~\onlinecite{Goldstein}.
The comparison to NRG is nevertheless much more satisfactory than with the RWA, 
as seen in the lower panel of Fig.~\ref{CompareRWA}. The only discrepancy comes again from the 
lack of inelastic contributions within the NIBA expression.

\begin{figure}[tb]
\includegraphics[width=0.99\columnwidth]{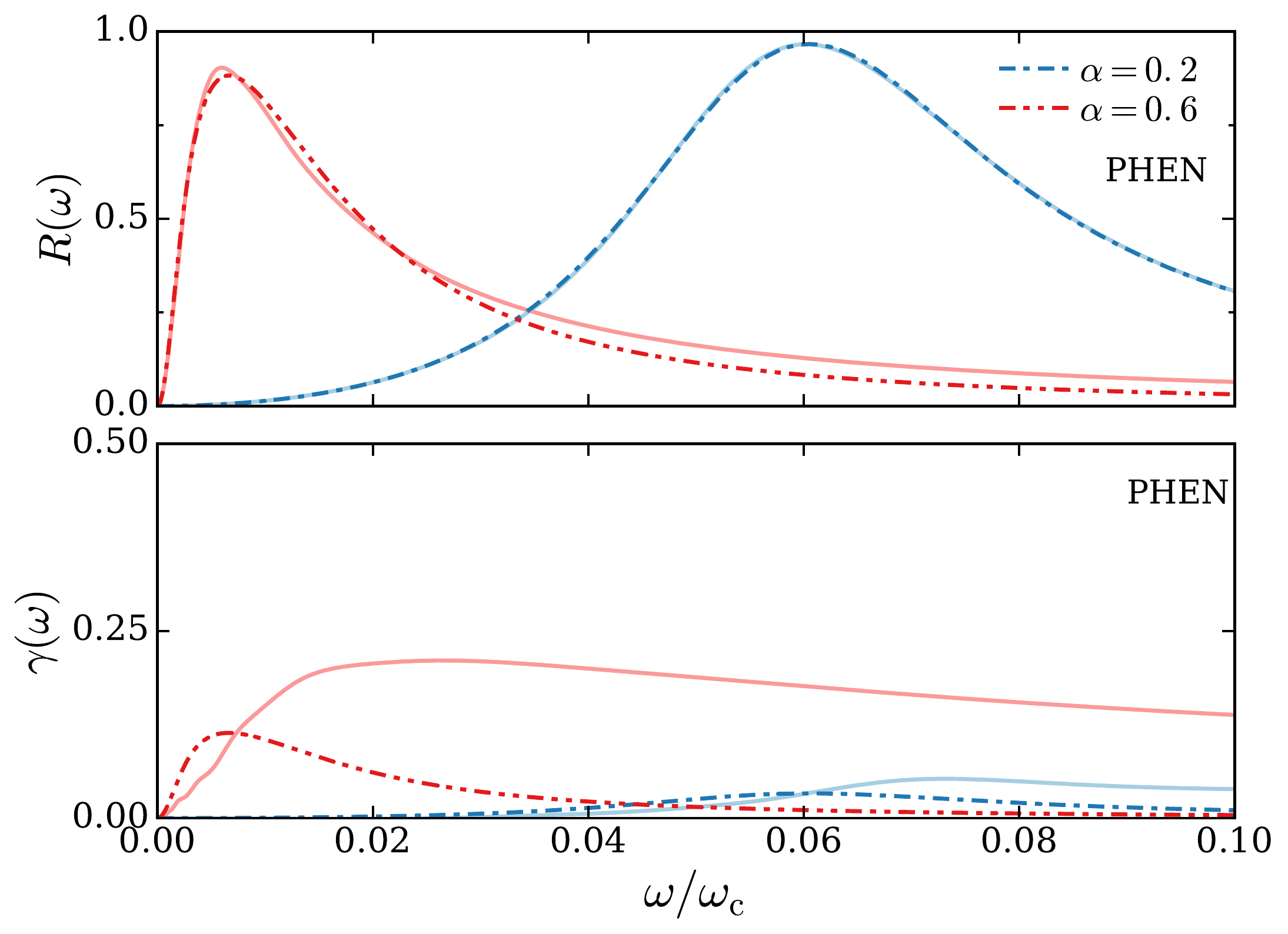}
\caption{(Color online) Reflection coefficient (top panel) and inelastic deficit
(bottom panel) comparing the phenomenological lineshape
Eq.~(\ref{RPHEN}) (dashed lines) to the exact NRG 
calculations (full lines) at two values of dissipation $\alpha=0.2,0.6$
(rightmost and leftmost curves respectively).
An effective renormalized qubit splitting $\Delta_R$ and a renormalized linewidth 
$\Gamma$ were used as fitting parameters within the phenomenological formula, for better 
comparison of the actual lineshapes.}
\label{ComparePHEN}
\end{figure}

The NIBA expression can, in fact, be corrected phenomenologically by noticing 
a simple problem in the formula for the spin susceptibility~(\ref{chiNIBA}).
Indeed, the sum rule~(\ref{sumrule}) is increasingly violated as $\alpha$
grows:
\begin{eqnarray}
\nonumber
\int_{-\infty}^{+\infty} \!\! \mr{d}\w \; \chi_\mr{NIBA}''(\w)
\mr{Sign}(\w) 
&=& \frac{\pi+2\,\mr{atan}\left[\frac{2-(2\pi\alpha)^2}
{2\pi\alpha\sqrt{4-(2\pi\alpha)^2}}\right]}{2\sqrt{4-(2\pi\alpha)^2}}\\
&\equiv&\frac{\pi}{2} N_\alpha.
\end{eqnarray}
One recovers $N_\alpha\to 1$ for $\alpha\to0$, but in general the deviation from 
the sum rule can be large, and, as we will see, accounts for the missing inelastic contribution.

We thus define an improved phenomenological formula:
\begin{equation}
\label{chiPhen}
\chi_\mr{phen.}(\w) = \frac{1}{N_\alpha}\frac{\Delta_R}{\Delta_R^2-\w^2 - i
2\pi\alpha\w\Delta_R},
\end{equation}
which satisfies by construction the exact sum rule. This leads to the following
expressions for the reflection, transmission, and inelastic responses:
\begin{eqnarray}
\label{RPHEN}
R_\mr{phen.} &=& \frac{1}{N_\alpha^2}
\frac{(2\Gamma\w)^2}{(\Delta_R^2-\w^2)^2+(2\Gamma\w)^2}\\
T_\mr{phen.} &=& 
\frac{(\Delta_R^2-\w^2)^2+(1-N_\alpha^{-1})^2 (2\Gamma\w)^2}
{(\Delta_R^2-\w^2)^2+(2\Gamma\w)^2}\\
\gamma_\mr{phen.} &=& 
\frac{N_\alpha-1}{N_\alpha^2} \frac{2 (2\Gamma\w)^2}
{(\Delta_R^2-\w^2)^2+(2\Gamma\w)^2}.
\end{eqnarray}
The comparison to the NRG data in the top panel of Fig.~\ref{ComparePHEN} gives now 
excellent agreement at $\alpha\leq0.2$ for the reflection coefficient, and reproduces 
well the magnitude of the inelastic deficit, although the shift of the peak position
and the long tails in $\gamma(\w)$ are not correctly accounted for. 
We also see from the bottom panel of Fig.~\ref{ComparePHEN} that large deviations 
occur in $\gamma(\w)$ for larger $\alpha$ values; this is expected~\cite{FlorensFritz} 
because a slower decay of the form $\chi''(\w)\sim 1/(\w\log^2\w)$ is known to occur in the Kondo 
regime for $\w\gg\Delta_R$. As a result, the inelastic contribution is 
even more underestimated in this regime.

\section{Scattering properties: Coherent state dynamics}
\label{Scattering}

\subsection{Coherent state formalism for a two-terminal waveguide}

The problem we want to study is the scattering of a coherent state
wavepacket off the Silbey-Harris ground state of the spin-boson model.
The strategy is to define the proper initial condition for the wavepacket
in left-right space (the full waveguide), then transform it to an 
even-odd basis, and propagate the resulting state. Since only even modes
couple to the qubit, the equations of motion~(\ref{eqf})-(\ref{eqq}) can
be used, while the odd modes are freely propagating. After the scattering
process has occured, a transformation back to the left-right basis 
provides the final wavefunction, which allows one to find the
transmitted and reflected intensity. 

Mathematically, we denote the initial scattering wavepacket by 
$z_{x}$ or $z_{k}$ in position space or momentum space, respectively 
(this excludes the 
entanglement cloud associated to the Silbey-Harris ground state). These
are related by the Fourier transform convention,
\begin{equation}
z_{x}=\int_{-\infty}^{\infty}\frac{dk}{\sqrt{2\pi}}e^{+ikx}z_{k}.
\end{equation}
The even and odd parts of the wavepaket are then defined strictly for $k>0$ as
\begin{equation}
z_{k}^{e}=\frac{1}{\sqrt{2}}\left(z_{k}+z_{-k}\right)\quad\textrm{and}\quad
z_{k}^{o}=\frac{1}{\sqrt{2}}\left(z_{k}-z_{-k}\right),
\end{equation}
which can be inverted to yield
\begin{equation}
z_{k}=\frac{1}{\sqrt{2}}\left(z_{k}^{e}+z_{k}^{o}\right)\quad\textrm{and}\quad
z_{-k}=\frac{1}{\sqrt{2}}\left(z_{k}^{e}-z_{k}^{o}\right).
\label{wavepacketbasis}
\end{equation}
Note that the first expression in Eq.~(\ref{wavepacketbasis}) gives the 
\emph{right-going} wave ($k>0$), while the second one is the \emph{left-going}
wave ($-k<0$).

We choose to use a Gaussian wavepacket defined as
\begin{equation}
z_{k}=\sqrt{\bar n} \left(\frac{1}{2\pi\sigma^{2}}\right)^{\frac{1}{4}}e^{-\frac{(k-k_{0})^{2}}{4\sigma^{2}}}e^{-i(k-k_{0})x_{0}}e^{-ik_{0}x_{0}/2},
\end{equation}
corresponding to a signal initially centered around $x_{0}$, with mean
wavenumber $k_{0}$, spatial extent $1/\sigma$, and a total intensity
corresponding to $\bar{n}$ photons on average.
The corresponding real space wavepacket is then
\begin{equation}
z_{x}=\sqrt{\bar n} \left(\frac{2\sigma^{2}}{\pi}\right)^{\frac{1}{4}}e^{-(x-x_{0})^{2}\sigma^{2}}e^{+ik_{0}(x-x_{0})}e^{+ik_{0}x_{0}/2}.
\end{equation}
Note that these amplitudes are both normalized so that
$\int_{-\infty}^{\infty}dx|z_{x}|^{2}=\int_{-\infty}^{\infty}dk|z_{k}|^{2}
=\bar n$.

Now let us define the actual photon content of the wavepacket, using here
coherent states, which are better suited for our simulations than scattering 
Fock states. 
The creation operator for a photon in the wavepacket state is defined as
\begin{equation}
a_{z}^{\dagger}=\int_{-\infty}^{+\infty}dk\,
z_{k\,}a_{k}^{\dagger}=\int_{-\infty}^{+\infty}dx\, z_{x\,}a_{x}^{\dagger}.
\end{equation}
A coherent state with an average of $\bar{n}$ photons in this wavepacket
is thus:
\begin{equation}
e^{-\bar{n}/2}e^{a_{z}^{\dagger}}\left|0\right\rangle =e^{a_{z}^{\dagger}-a_{z}}\left|0\right\rangle .
\end{equation}
One can construct even and odd creation and destruction operators for
bosons in a given mode $k>0$ by analogy to the transformations made previously on the wavepacket:
\begin{equation}
a_{k}^{e}=\frac{1}{\sqrt{2}}\left(a_{k}+a_{-k}\right)\quad\textrm{and}\quad
a_{k}^{o}=\frac{1}{\sqrt{2}}\left(a_{k}-a_{-k}\right).
\end{equation}
Then the creation operator for the wavepacket state is
\begin{eqnarray}
a_{z}^{\dagger} &=&
\int_{0}^{+\infty}dk\, \left( z_{k\,}a_{k}^{\dagger}+ z_{-k\,}a_{-k}^{\dagger}\right)\\
&=&
\int_{0}^{+\infty}dk\,\left(z_{k\,}\frac{a_{k}^{e\dagger}+a_{k}^{o\dagger}}{\sqrt{2}}+z_{-k\,}\frac{a_{k}^{e\dagger}-a_{k}^{o\dagger}}{\sqrt{2}}\right)\\
&=&
\int_{0}^{+\infty}dk\,\left(z_{k\,}^{e}a_{k}^{e\dagger}+z_{k\,}^{o}a_{k}^{o\dagger}\right)
= a_{z}^{e\dagger}+a_{z}^{o\dagger}.
\end{eqnarray}
Notice that the even and odd sector operators commute,
\begin{equation}
[a_{k}^{e\dagger},a_{k}^{o}]=[a_{z}^{e\dagger},a_{z}^{o}]=0.
\end{equation}
Thus, the coherent state wavepacket can finally be written as:
\begin{equation}
e^{-\bar{n}/2}e^{a_{z}^{\dagger}}\left|0\right\rangle
=e^{a_{z}^{o\dagger}}e^{-\bar{n}/2}e^{a_{z}^{e\dagger}}\left|0\right\rangle
=e^{a_{z}^{o\dagger}}e^{a_{z}^{e\dagger}-a_{z}^{e}}\left|0\right\rangle .
\end{equation}

The final step in the initialization of the state vector is to combine 
the scattering wavepacket coherent state $z_k$ with the Silbey-Harris ground
state with displacement $f_k^\mr{SH}=(1/2) g_k/(\w_k+\Delta_R)$. 
Since the even and odd sectors are completely independent, the Silbey-Harris
state affects only the even sector. 
For the spin-up projection of the wavefunction, we have thus,
\begin{eqnarray}
e^{a_{z}^{e\dagger}-a_{z}^{e}}\left|\Psi_\ua\right\rangle 
&=&e^{\sum_{k>0}\left(z_{k}^{e}a_{k}^{e\dagger}-z_{k}^{e*}a_{k}^{e}\right)}\\
\nonumber
&& \times \, e^{\sum_{k>0}\left(f_{k}^\mr{SH}a_{k}^{e\dagger}-f_{k}^{\mr{SH}*}a_{k}^{e}\right)}
\left|0\right\rangle ,
\end{eqnarray}
where for compactness we have switched to sums over $k$ instead of integrals. 
The two exponentials can be combined, keeping in mind that there may be a
phase from 
the commutator in the standard relation
$ e^{A}e^{B}=e^{A+B}e^{\frac{1}{2}[A,B]}$, which is valid since the commutator
here is just a $c$-number.
Thus the initial state in the $\ua$ projection is
\begin{eqnarray}
e^{a_{z}^{e\dagger}-a_{z}^{e}}\left|\Psi_\ua\right\rangle 
&=&e^{\frac{1}{2}\sum_{k>0}\left(z_{k}^{e}f_{k}^{\mr{SH}*}-z_{k}^{e*}f_{k}^\mr{SH}\right)}\\
\nonumber
&& \times \, e^{\sum_{k>0}\left[\left(f_{k}^\mr{SH}+z_{k}^{e}\right)
a_{k}^{e\dagger}-\left(f_{k}^\mr{SH}+z_{k}^{e}\right)^{*}a_{k}^{e}\right]}\left|0\right\rangle .
\end{eqnarray}
For the spin-down projection, one simply replaces $f_{k}^\mr{SH}$
by $-f_{k}^\mr{SH}$ without changing the sign of $z_{k}^{e}$, 
so that our total initial wavefunction reads:
\begin{eqnarray}
\left|\Psi(t=0)\right\rangle 
&=&\frac{1}{\sqrt{2}} \ket{\ua} 
e^{\frac{1}{2}\sum_{k>0}\left(z_{k}^{e}f_{k}^{\mr{SH}*}-z_{k}^{e*}f_{k}^\mr{SH}\right)}\\
\nonumber
&&\times\, e^{\sum_{k>0}\left[\left(f_{k}^\mr{SH}+z_{k}^{e}\right)
a_{k}^{e\dagger}-\left(f_{k}^\mr{SH}+z_{k}^{e}\right)^{*}a_{k}^{e}\right]}\left|0\right\rangle
\\
\nonumber
&&- \frac{1}{\sqrt{2}} \ket{\da} 
e^{\frac{1}{2}\sum_{k>0}\left(-z_{k}^{e}f_{k}^{\mr{SH}*}+z_{k}^{e*}f_{k}^\mr{SH}\right)}\\
\nonumber
&&\times\,
e^{\sum_{k>0}\left[\left(-f_{k}^\mr{SH}+z_{k}^{e}\right) a_{k}^{e\dagger}
-\left(-f_{k}^\mr{SH}+z_{k}^{e}\right)^{*}a_{k}^{e}\right]}\left|0\right\rangle.
\end{eqnarray}
One can thus use this state as initial condition for the equations of
motion~(\ref{eqf})-(\ref{eqq}), with $f_k(0)=f_k^\mr{SH}+z_k^e$,
$h_k(0)=-f_k^\mr{SH}+z_k^e$, and $p(0)=-q(0)=1/\sqrt{2}$.
Note that the odd sector, which decouples from the qubit, is trivially evolving
in time according to $i \dot{z}_k^o = \w_k z_k^o$.

After this time-evolution is performed, the output must be analyzed.
Because of the constrained ansatz~(\ref{Psi}), the output is again a 
single coherent state, but one has to recombine the even sector output with 
the odd sector trivial evolution. After a long enough time $T$ so that the 
outgoing wavepacket is disentangled from the Silbey-Harris cloud, one obtains 
the total displacement $f_k(T) = f_k^\mr{SH} + z_{k,\mr{out}}^e$, separating
into the ground state contribution and an outgoing scattering part.
Combining with $z_k^o(T) = e^{-i\w_k t} z_k^o(0)$ using 
Eq.~(\ref{wavepacketbasis}), this provides the final outgoing coherent
state displacement $z_{k,\mr{out}}$ in the physical momentum basis.
One can then define transmission and reflection coefficients as the ratio of the
outgoing (in each possible terminal) to the ingoing power:
\begin{equation}
T=\frac{\sum_{k>0}k\left|z_{k,\textrm{out}}\right|^{2}}{\sum_{k>0}k\left|z_{k,i\textrm{n}}\right|^{2}}\quad\textrm{and}\quad
R=\frac{\sum_{k<0}|k|\left|z_{k,\textrm{out}}\right|^{2}}
{\sum_{k>0}k\left|z_{k,i\textrm{n}}\right|^{2}}.
\end{equation}

\subsection{Comparison of NRG and coherent state dynamics results for the
transport coefficients}

\begin{figure}[tb]
\includegraphics[width=1.0\columnwidth]{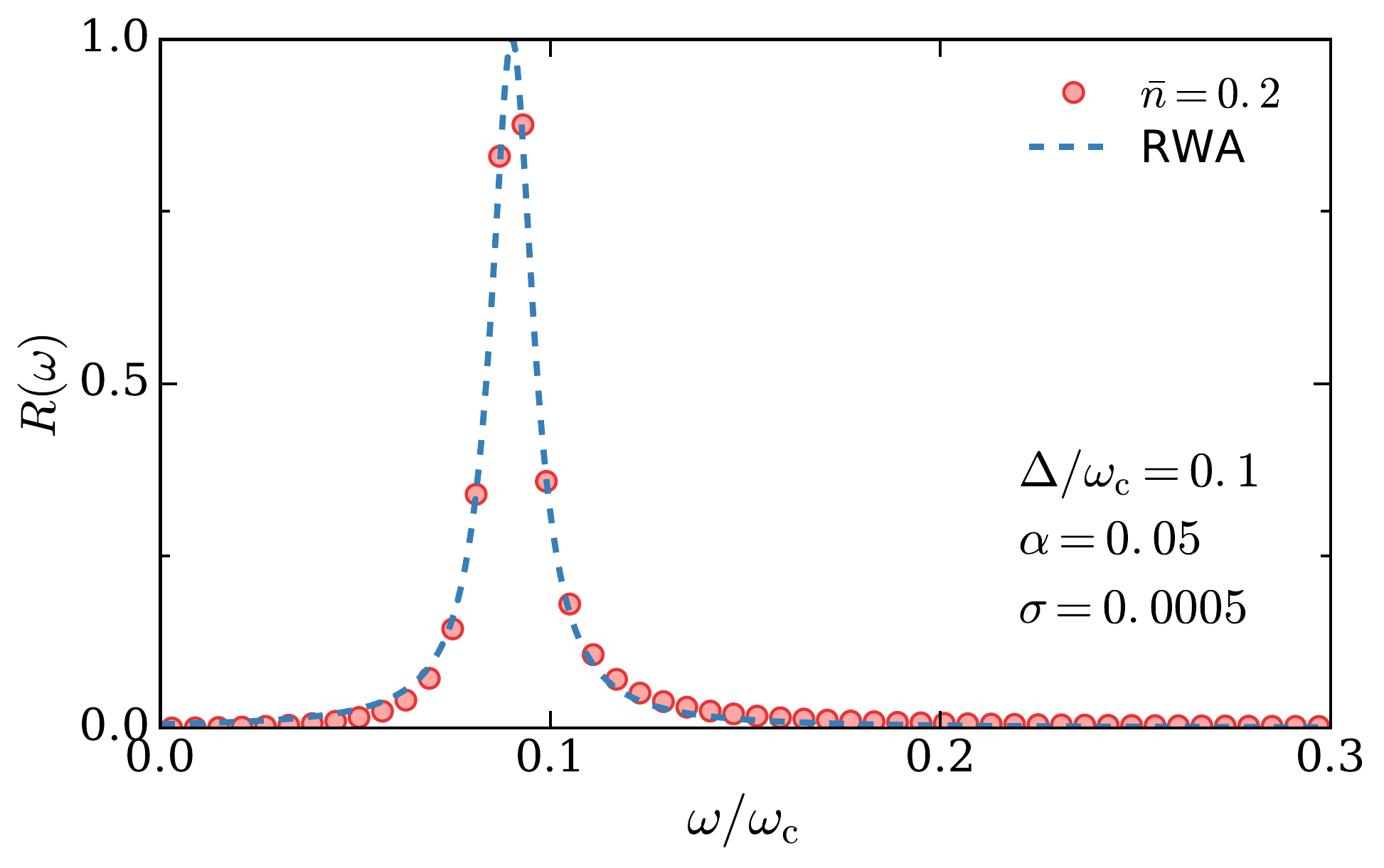}
\caption{(Color online) 
In the quantum optics regime $\alpha=0.05$, the reflection coefficient obtained 
from the variational dynamics (symbols) matches
the RWA lineshape (dashed line), which was however corrected with the proper
renormalized qubit splitting. Parameters here are $\Delta/\w_c=0.1$
and the variational dynamics is done at weak power ($\bar{n}=0.2$) in order to be in the
linear response regime. The wavepacket width is taken very small $\sigma=0.0005$ to 
allow good spectral resolution.}
\label{CoherentStateTransmissionSmallAlpha}
\end{figure}

\begin{figure}[tb]
\includegraphics[width=1.0\columnwidth]{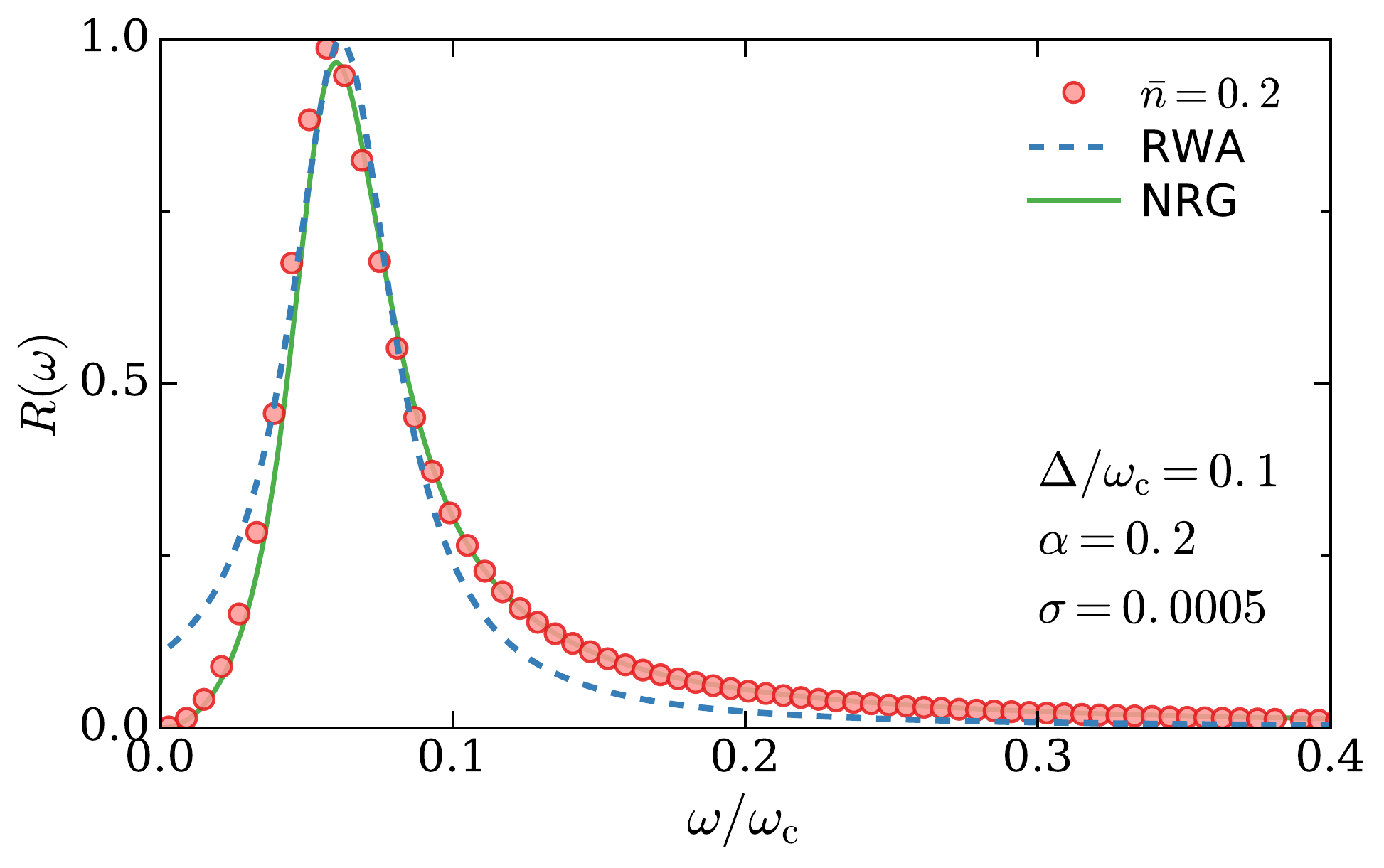}
\caption{(Color online) 
In the ultrastrong coupling regime $\alpha=0.2$, the reflection
coefficient obtained from the variational dynamics (symbols) 
compares favorably to the exact NRG lineshape (solid line), while 
the RWA (dashed line) presents clear deviations. 
The same parameters as in Fig.~\ref{CoherentStateTransmissionSmallAlpha} were used.}
\label{CoherentStateTransmissionLargeAlpha}
\end{figure}

In this last section, we present transport calculations using the
single coherent state dynamics and benchmark them
in the linear response regime against the controlled results from NRG,
including a comparison to RWA as well.
Let us first consider the quantum optics regime $\alpha\ll1$ 
where the RWA can be trusted,
as shown in Fig.~\ref{CoherentStateTransmissionSmallAlpha}
for $\alpha=0.05$ and an incident coherent state wavepacket whose intensity is small.
We see excellent agreement between the two methods; thus, the coherent
state approach performs very well in the quantum optics regime [note
that an effective qubit splitting was added by hand in the RWA curve].

Increasing dissipation up to $\alpha=0.2$, 
we assess the validity of both the coherent state dynamics and the RWA by comparing to the numerically exact NRG results in Fig.~\ref{CoherentStateTransmissionLargeAlpha}. While the RWA results are rather inaccurate, the coherent state dynamics is in excellent agreement with the NRG, showing its utility well beyond the usual quantum optics regime. 

\begin{figure}[tb]
\includegraphics[width=1.0\columnwidth]{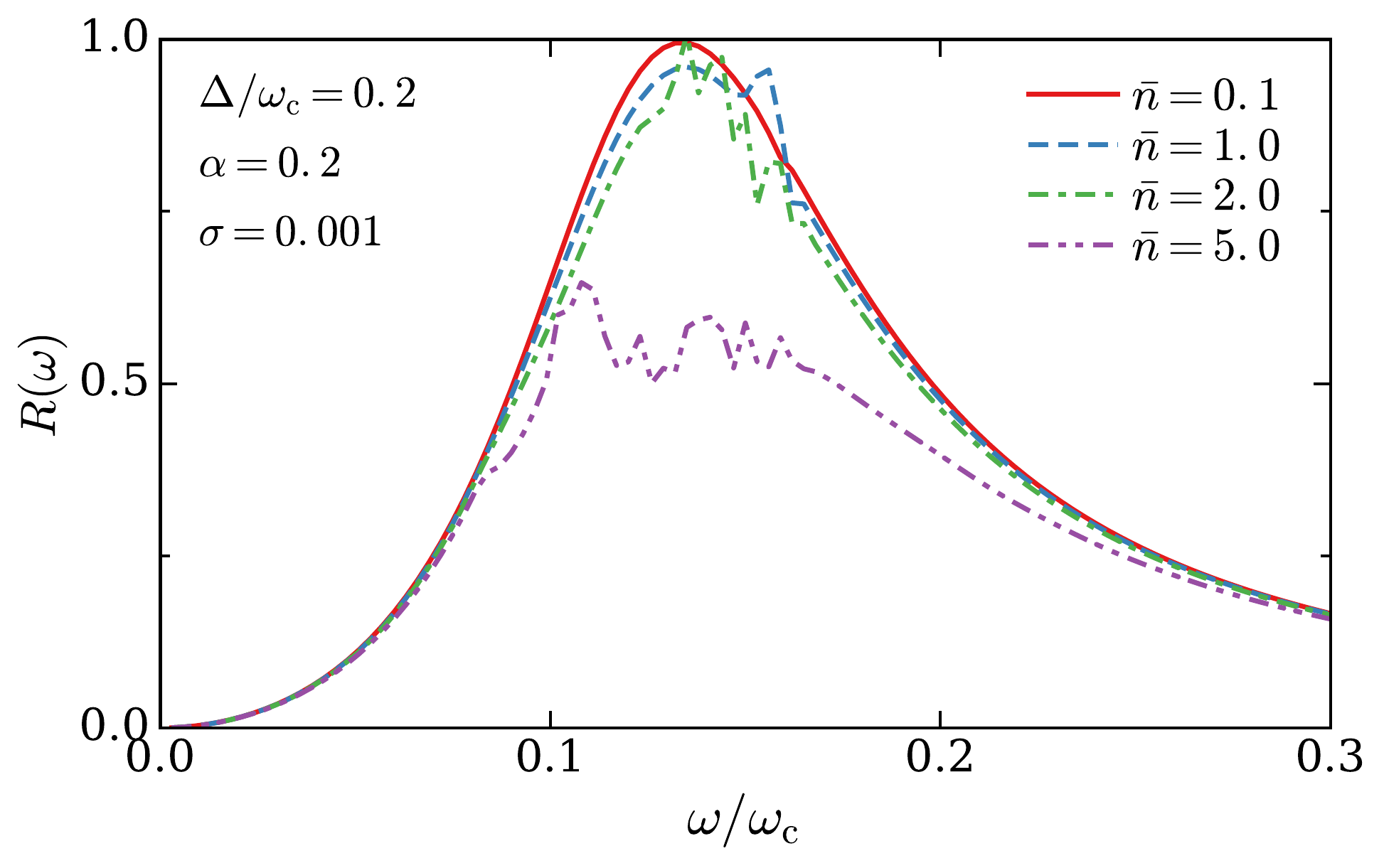}
\caption{(Color online) Reflection coefficient at $\alpha=0.2$
and $\Delta/\w_c=0.2$ for increasing irradiation power. 
The incident power is characterized by the average number of photons $\bar{n}$ in 
the incoming coherent state wavepackets; results for up to $\bar{n}=5$ are shown.}
\label{power}
\end{figure}
  
Being a non-equilibrium method, the coherent state dynamics can access regimes
beyond linear response, which the present formulation of the NRG is not able
to achieve. We show for instance in Fig.~\ref{power} how the transport
coefficients evolve as input power is turned on, increasing the average number 
of photons in the incoming coherent state wavepacket up to $\bar{n}=5$.
Having checked our numerical integration carefully,
we attribute the large level of noise in these results to the restricted form of
the ansatz~(\ref{Psi}) and not to numerical inaccuracies. This is
consistent with our general observation that artifacts typically arise for 
strong temporal perturbations in the single coherent state dynamics, which
indeed increase at larger power.

The general trend of the curves are however clear~\cite{LeHur}: due to saturation effects
of the two-level system by the large flux of photons, the reflection is
reduced at increasing power.
Clear deviations from the linear response result occur for an average
photon number $\bar{n}\simeq 5$. This value can be understood from the fact that the qubit
excitation rate $\bar{n}\sigma=5 \times 0.005=0.025$ becomes then comparable to the qubit
decay rate $\Gamma\simeq \alpha \Delta_R = 0.2\times 0.15\simeq 0.03$.
Saturation effects are indeed expected to be governed by the renormalized qubit
frequency at strong coupling.

\section{Conclusion}

We have investigated here the relaxation and transport properties of a single
qubit side-coupled to a large impedance transmission line, using complementary
many-body methods. The numerical renormalization group (NRG) allowed us to
accurately compute the linear-response transport coefficients, and showed good 
agreement for a simpler approach using the variational 
time-evolution of a wavefunction based on a single coherent state. This coherent
state dynamics was benchmarked also regarding relaxation properties, and despite
some success in predicting the onset of overdamped dynamics, showed some
physical artifacts such as improper relaxation to the correct steady state.
We plan to overcome these difficulties in future works by systematically extending the
dynamical ansatz along the lines of Ref.~\cite{Bera2}, allowing us also to tackle the transport problem in a more controlled way.

\acknowledgments
We thank Zach Blunden-Codd, Alex Chin, Ahsan Nazir, Nicolas Roch, Marco 
Schir\'o and Izak Snyman for stimulating discussions and acknowledge funding 
from the Fondation Nanosciences de Grenoble under RTRA contract CORTRANO. 
The work at Duke was supported by the U.S.\ DOE, Division of 
Materials Sciences and Engineering, under Grant No.\ DE-SC0005237.

\end{document}